\documentstyle[aps, preprint]{revtex}
\begin{document}
\draft
\preprint{gr-qc/9907032}
\tightenlines

\title{Wave Functions for Quantum Black Hole Formation in
Scalar Field Collapse}
\author{Dongsu
Bak\footnote{Electronic address: dsbak@mach.uos.ac.kr}$^{a}$, Sang
Pyo Kim\footnote{Electronic address:
sangkim@knusun1.kunsan.ac.kr}$^{b}$, Sung Ku
Kim\footnote{Electronic address: skkim@theory.ewha.ac.kr}$^{c}$,
Kwang-Sup Soh\footnote{Electronic address:
kssoh@phya.snu.ac.kr}$^{d}$, and Jae Hyung Yee\footnote{Electronic
address: jhyee@phya.yonsei.ac.kr}$^{e}$}
\address{a Department of Physics, University of Seoul, Seoul
130-743 Korea\\
b Department of Physics, Kunsan National University, Kunsan
573-701 Korea\\
c Department of Physics, Ewha Women's University, Seoul 120-750
Korea\\
d Department of Physics Education, Seoul National University,
Seoul 151-742 Korea\\
e Department of Physics, Yonsei University, Seoul 120-749 Korea}

\date{\today}

\maketitle

\begin{abstract}
We study quantum  mechanically the self-similar  black hole
formation by   collapsing scalar field and  find the  wave
functions that  give the  correct semiclassical  limit. In
contrast to classical theory, the wave functions for the black
hole formation even in  the supercritical case have not only
incoming flux but also outgoing flux.   From this result we
compute  the rate  for the   black hole formation.
In the subcritical case our result agrees with the semiclassical tunneling rate.
Furthermore, we  show  how to recover the  classical evolution  of  black hole
formation from  the  wave function  by defining the
Hamilton-Jacobi characteristic function as  $W = \hbar {\rm Im}
\ln \psi$. We find that the quantum  corrected apparent horizon
deviates from  the classical value
only slightly without any qualitative change even in the critical case.
\end{abstract}

\section{Introduction}

Matters can strongly interact gravitationally to form black holes.
The Schwarzschild, Reissner-Nordtr\"{o}m, Kerr and Kerr-Newman
black holes are supposed to describe the end state of such
gravitationally collapsing objects and are characterized by mass,
charge, and angular momentum. Since these black holes are
stationary, it would be physically interesting to understand the
dynamical process for black hole formation that leads to the
stationary black hole as the end state. Such an attempt toward
understanding the black hole formation process was first done by
Christodoulou who analytically studied a massless scalar field to
prove that it either forms a black hole for the strong
self-gravitational interaction or disperses for the weak
self-gravitational interaction \cite{christodoulou}. The
gravitational collapse of the massless scalar field was confirmed
in an early numerical simulation \cite{goldwirth}. The most
important results of gravitational collapse was first discovered
by Choptuik who found through numerical investigation that an
initial Gaussian packet for the massless scalar field collapses
self-similarly from spatial infinity either to form a black hole
or to disperse back into spatial infinity depending on whether the
parameter characterizing the wave packet is above or below a
critical value, and that the black hole mass exhibits a power-law
scaling behavior whose critical exponent is independent of the
initial data \cite{choptuik}. Since then critical phenomena of
gravitational collapse in the spherically symmetric geometry have
been found numerically in the perfect fluid
\cite{evans,koike,maison}, complex scalar
\cite{hirschmann,gundlach,hod}, SU(2) Einstein-Yang-Mills
\cite{gundlach2}, non-linear $\sigma$ \cite{hirschmann2},
axion-dilaton \cite{eardley,hamade}, and gravitational wave models
\cite{abrahams}.

The simplest model for the gravitational collapse is a massless
scalar field evolving self-similarly in the spherical-symmetric
geometry, and has been studied analytically \cite{christodoulou}
and numerically \cite{goldwirth,choptuik}. The analytic solutions
\cite{roberts} for that model together with the self-similarity
has been used to study the critical behavior
\cite{brady,frolov,oshiro}. Most of these works treat the
classical aspects of gravitational collapse. Since the massless
scalar field model predicts the type-II critical behavior, the
mass of black hole near the critical parameter can be arbitrarily
small. So one may ask whether quantum effects play a significant
role in the system. Hence it would be physically interesting and
important to study the quantum gravitational collapse. As a
quantum mechanical treatment of gravitational collapse, the wave
function of a supercritical parameter for a quantum black hole was
expressed in terms those of the subcritical parameters and thus
explains the black hole decay quantum mechanically
\cite{tomimatsu}. Recently, we have also studied such quantum
effects that, though classically the collapsing scalar field below
the critical parameter does not form a black hole, quantum
mechanically a black hole can be formed through tunneling process
\cite{bak}. However, the interpretation of the wave functions in
the supercritical and subcritical cases has not completely been
settled yet \cite{tomimatsu,bak}, and a question has been raised
quite recently how quantum effects can change the physics such as
the black hole mass or critical exponent near the critical value
\cite{choptuik2}.

It is the purpose of this paper to study quantum mechanically the
black hole formation in the massless scalar field model and to
investigate how quantum effects modify the classical picture of
gravitational collapse. We use the ADM formulation to quantize the
model and find analytically the black hole wave function in terms
of the confluent hypergeometric function. The quantum model
resembles in many respects a quantum FRW cosmological model
minimally coupled to a massless scalar field. The standard
interpretation of quantum mechanics and quantum cosmology enables
us to calculate explicitly the rate for black hole formation. We
put forth a quantum criterion on the black hole formation, whose
semiclassical approximation agrees with the classical one.
Furthermore, by using the characteristic function defined by $W =
\hbar{\rm Im} \ln \psi$ we are able to compute the quantum effects
to the classical apparent horizon and to the critical exponent.

The organization of this paper is as follows. In Sec. II, using
the ADM formulation we derive  the Wheeler-DeWitt equation for
black hole formation by a self-similarly collapsing  scalar field.
The Wheeler-DeWitt equation has  an SU(2)  group structure and the
wave functions are found in terms of the confluent hypergeometric
function. In Sec. III, we find the wave function for the black
hole formation, which consists of both the incoming and outgoing
components. We calculate the incident and transmitted fluxes, in
terms of which the rate for black hole formation is found. It is
found that the rate for black hole formation gives the correct
semiclassical limits in  both the supercritical and subcritical
cases. In Sec. IV we use the new method for recovering the
classical solution from our wave function introduced by us
recently  in Ref. \cite{bak2}. The  key idea is to  modify the
Wheeler-DeWitt equation to have a nonvanishing energy $E$ rather
than the usual vanishing value. Then we define the Hamilton-Jacobi
characteristic function  as $W = \hbar  {\rm Im} \ln \psi$,  from
which we  get the Hamilton-Jacobi  evolution of classical
solution. Finally we let the energy $E$ vanish. Following this
procedure we regain the semiclassical limit of our quantum
solutions and find the position of the apparent horizon. We have
used the steepest descent method to determine the position of
apparent horizon and found good agreement with numerical
evaluation. Even in the critical case ($c_0 = 1$) there appears no
qualitative change from the classical results. In the last
section, we discuss the salient points of our work.

\section{Canonical Quantization}

The model to be studied in this paper is the spherical-symmetric
geometry minimally coupled to a massless scalar field. Then the
Hilbert-Einstein action
\begin{equation}
 S = \frac{1}{16\pi} \int_{M}d^{4}x \sqrt{-g}~\Biggl[\frac{1}{G}
 {}^{(4)}R - 2(\nabla \phi)^{2} \Biggr] + {1\over 8\pi G} \int_{\partial M}
 d^3x K \sqrt{h},  \label{act1}
\end{equation}
where $G$ is the Newton constant, reduces to the $(1+1)$-action
\begin{equation}
S_{sph} = \frac{1}{4} \int \ d^{2}x \sqrt{-\gamma} ~r^{2}
\Biggl[\frac{1}{G} \Bigl\{{}^{(2)} R(\gamma)+\frac{2}{r^{2}}
\Bigl( (\nabla r)^{2} +1 \Bigr) \Bigr\} -2(\nabla \phi)^{2}
\Biggr], \label{act2}
\end{equation}
where $\gamma_{ab}$ is the metric in the remaining two-dimensional
manifold. The $(1+1)$-metric for the scalar field collapse has the
form
\begin{equation}
{}^{(2)} ds^2 = - 2 e^{2 \sigma} du dv
\end{equation}
where $u$ and $v$ are null coordinates, and $\sigma$ is a function
of $u$ and $v$. To find the continuously self-similar solutions,
$r = - u \sqrt{z} y = \sqrt{- uv} y$, $\phi(z)$ and $\sigma (z)$,
which depend only on $z = - \frac{v}{u} = e^{-2\tau}$, we
introduce a diagonal gauge
\begin{equation}
u = - \omega e^{- \tau},~~ v = \omega e^{\tau},
\end{equation}
in terms of which the $(1+1)$-metric takes the form
\begin{equation}
{}^{(2)} ds^2 = e^{2 \sigma} \Bigl( - 2 \omega^2 d \tau^2 + 2
d\omega^2 \Bigr).
\end{equation}
The classical solutions with vanishing $\sigma$ are found in Refs.
\cite{brady,frolov}, and the corresponding spacetime has the
geometry ${\cal M} = R^{1,1} \times S^2$.

In order to canonically quantize the system we adopt the ADM
formulation and introduce a lapse function $N(\tau)$ to write the
$(1+1)$-metric as
\begin{equation}
ds^2_{(2)} = - 2 N^2 (\tau) \omega^2 d\tau^2 + 2 d\omega^2.
\end{equation}
After taking into account the boundary terms, we obtain the
action\footnote{In this paper we use a specific unit system $c =
1$ keeping explicitly the Planck constant $\hbar$ and the
gravitational constant $G$, in which $m_P =
\sqrt{\frac{\hbar}{G}}, l_P = \sqrt{\hbar G}$, and $m_P =
\frac{\hbar}{l_P}$. The variables $\omega$ and $r$ have the
dimension of $l_P$ and the time-like variable $\tau$ and $y$
become dimensionless.}
\begin{equation}
S_{sph} = \int d\Bigl(\frac{m_P^2}{\hbar} \frac{\omega^2}{2}
\Bigr) d\tau \Biggl[-\frac{1}{2N} \dot{y}^{2}+\frac{\hbar
y^2}{2m_P^2 N} \dot{\phi}^{2} - N \Bigl\{ \frac{1}{2}y^2 - 1 - 2
\Bigl(\omega\frac{\partial}{\partial\omega}y \Bigr)^{2} +
\frac{2\hbar}{m_P^2} \Bigl(
\omega\frac{\partial}{\partial\omega}\phi
\Bigr)^{2}\Bigr\}\Biggr], \label{act3}
\end{equation}
where overdots denote derivative with respect to $\tau$. Note that
$K \equiv \frac{m_P^2}{\hbar} \frac{\omega^2}{2} = \hbar
\frac{\omega^2}{2 l_P^2}$ has the dimension of action $\hbar$ and
is proportional to the area factor measured in the Planck length
$l_P$, and plays the role of a cut-off parameter of the model.
Hence $\frac{K}{\hbar}$ is a dimensionless parameter. The
canonical momenta for $y$ and $\phi$ are given by
\begin{equation}
\pi_{y} = - \frac{K}{N} \dot{y}, ~~\pi_{\phi} = \frac{\hbar K
y^2}{m_P^2 N} \dot{\phi}.
\end{equation}
Then Eq. (\ref{act3}) can be rewritten in the ADM formulation as
\begin{equation}
S_{sph} = \int dx \Biggl[ \pi_{y} \dot{y} + \pi_{\phi} \dot{\phi}
- N {\cal H} \Biggr],
\end{equation}
where
\begin{equation}
{\cal H} = - \frac{1}{2K} \pi_y^2 +\frac{m_P^2}{2 \hbar Ky^2}
\pi_{\phi}^2 - K \Bigl\{1 - \frac{1}{2}y^2 + 2
\Bigl(\omega\frac{\partial}{\partial\omega}y \Bigr)^{2} - \frac{2
\hbar }{m_P^2} \Bigl(\omega\frac{\partial}{\partial\omega}\phi
\Bigr)^{2}\Bigr\},
\end{equation}
is the Hamiltonian. The $N$ acts as a Lagrange multiplier, so one
gets the Hamiltonian constraint
\begin{equation}
{\cal H} = 0. \label{const}
\end{equation}
According to the Dirac quantization method, the constraint
(\ref{const}) becomes a quantum constraint on the wave function
\begin{equation}
\hat{\cal H} \Bigl(y, \pi_{y}; \phi, \pi_{\phi} \Bigr) \vert \Psi
\rangle = 0. \label{quant const}
\end{equation}
We further impose the constraints from the self-similarity
\begin{equation}
\frac{\partial \hat{\phi}}{\partial \omega} \vert \Psi  \rangle =
0 = \frac{\partial \hat{y}}{\partial \omega} \vert \Psi  \rangle.
\end{equation}
Finally we obtain the Wheeler-DeWitt equation for the quantum
black hole formation
\begin{equation}
\Biggl[ \frac{\hbar^2}{2K} \frac{\partial^2}{\partial y^2} -
\frac{\hbar m_P^2}{2Ky^2} \frac{\partial^2}{\partial \phi^2} - K
\Bigl(1 - \frac{y^2}{2} \Bigr) \Biggr] \Psi(y, \phi) = 0,
\label{wd eq}
\end{equation}
which is the same equation that was used to calculate the
tunneling rate for black hole formation in the subcritical case
\cite{bak}.

The wave function can be factorized into the scalar and
gravitational field parts
\begin{equation}
\Psi(y, \phi) = \exp \Bigl( \pm i \frac{K c_0}{\hbar^{1/2} m_P}
\phi \Bigr) \psi(y). \label{fac wav}
\end{equation}
Here the wave function for the scalar field was chosen to yield
the classical momentum $\pi_{\phi} = \frac{\hbar K y^2
\dot{\phi}}{m_P^2 N} = \pm K c_0$, where $c_0$ is a dimensionless
parameter. Then Eq. (\ref{wd eq}) reduces to the gravitational
field equation
\begin{equation}
\Biggl[ \frac{\hbar^2}{2K} \frac{\partial^2}{\partial y^2} - K
\Bigl( 1 - \frac{y^2}{2}  - \frac{c_0^2}{2 y^2}\Bigr) \Biggr]
\psi(y) = 0. \label{wd eq2}
\end{equation}
It is worthy to note that Eq. (\ref{wd eq2}), as one-dimensional
quantum system, describes an inverted Calogero model with energy
$K$ and has the group structure SU(2) with the basis
\cite{calogero}
\begin{equation}
\hat{L}_{-} = \frac{\hat{\pi}_y^2}{2} - \frac{K^2 c_0^2}{2
\hat{y}^2}, ~~ \hat{L}_{0} = \frac{\hat{\pi}_y \hat{y} + \hat{y}
\hat{\pi}_y }{2},~~ \hat{L}_{+} = \frac{\hat{y}^2}{2}.
\end{equation}
Due to the group structure we are able to find the solutions to
Eq. (\ref{wd eq2}) in  terms of the confluent hypergeometric
function \cite{abramowitz}:
\begin{equation}
\psi_{I}(y) = D_{I}  \Biggl[ \exp \Bigl(- \frac{i}{2}
\frac{K}{\hbar} y^2 \Bigr) \Biggr] \Bigl(\frac{K}{\hbar} y^2
\Bigr)^{\mu_{-}} M (a_{-}, b_{-}, i \frac{K}{\hbar} y^2),
\label{wav1}
\end{equation}
where
\begin{eqnarray}
\mu_{-} &=& \frac{1}{4} - \frac{i}{2\hbar}Q, \nonumber\\ a_{-} &=&
\frac{1}{2} - \frac{i}{2\hbar} (Q + K), \nonumber\\ b_{-} &=& 1 -
\frac{i}{\hbar} Q,\label{parameter}
\end{eqnarray}
with
\begin{equation}
Q = \Bigl(K^2 c_0^2 - \frac{\hbar^2}{4} \Bigr)^{1/2}.
\end{equation}
The other independent solution is given by
\begin{equation}
\psi_{II}(y) = D_{II} \Biggl[ \exp \Bigl(- \frac{i}{2}
\frac{K}{\hbar} y^2 \Bigr) \Biggr] \Bigl(\frac{K}{\hbar} y^2
\Bigr)^{\mu_{-}} U (a_{-}, b_{-}, i \frac{K}{\hbar} y^2).
\label{wav2}
\end{equation}
However, the wave function (\ref{wav2}) is a linear combination of
$\psi_{I}$ and another solution
\begin{equation}
\psi_{III}(y) = D_{III} \Biggl[ \exp \Bigl(- \frac{i}{2}
\frac{K}{\hbar} y^2 \Bigr) \Biggr] \Bigl(\frac{K}{\hbar} y^2
\Bigr)^{\mu_{+}} M (a_{+}, b_{+}, i \frac{K}{\hbar} y^2),
\label{wav3}
\end{equation}
where
\begin{eqnarray}
\mu_{+} &=& \frac{1}{4} + \frac{i}{2\hbar}Q, \nonumber\\ a_{+} &=&
\frac{1}{2} + \frac{i}{2\hbar} (Q - K), \nonumber\\ b_{+} &=& 1 +
\frac{i}{\hbar} Q. \label{parameter2}
\end{eqnarray}
In the above equations $D$'s are constants, and $\mu$'s, $a$'s and
$b$'s are dimensionless parameters. Hence Eqs. (\ref{wav1}),
(\ref{wav2}) and (\ref{wav3}) are dimensionless functions.

\section{Wave Function for Black Hole Formation}

In the classical context, the massless scalar field imploding
self-similarly from spatial infinity with momentum above the
critical value collapses to form a black hole without leaving any
remnant and with momentum below the critical value it reflects
back to spatial infinity \cite{frolov}. However, in the quantum
context, we may follow the analogy to the scattering problem of
quantum mechanics. We prescribe the {\it boundary condition} for
the wave functions for the black hole formation such that they
should be incident from spatial infinity and some part of them be
reflected by the potential barrier back to spatial infinity but
the remaining part be transmitted toward the black hole
singularity inside their apparent horizons.

We now wish to calculate the quantum mechanical rate for black
hole formation. It is not difficult to see that Eq. (\ref{wav1})
has the asymptotic form\cite{abramowitz} at spatial infinity
\begin{equation}
\psi_{BH} = \psi_{I}(y) \cong  \tilde{D}_{I}
\Biggl[\frac{\Gamma(b_{-})}{\Gamma( a_{+}^*)}e^{i \pi a_{-}}
\Bigl(i \frac{K}{\hbar} y^2 \Bigr)^{\mu_{-} - a_{-}} e^{-
\frac{i}{2} \frac{K}{\hbar} y^2} +
\frac{\Gamma(b_{-})}{\Gamma(a_{-})} \Bigl(i \frac{K}{\hbar} y^2
\Bigr)^{\mu_{-} - a_{+}^*} e^{ \frac{i}{2} \frac{K}{\hbar} y^2}
\Biggr], \label{asym}
\end{equation}
where $\tilde{D}_{I} = D_{I} (-i)^{\mu_{-}}$. To show that Eq.
(\ref{wav1}) satisfies indeed the boundary condition for the black
hole formation, we note that the first term describes the incoming
component and the second term describes the outgoing (reflected)
component at spatial infinity, $( y \gg 1)$. Near $y = 0$, Eq.
(\ref{wav1}) has the form\cite{abramowitz}
\begin{equation}
\psi_{BH} \cong D_I  \Bigl( \frac{K}{\hbar}
y^2\Bigr)^{\frac{1}{4}} \exp \Bigl[-\frac{i}{2}
\Bigl(\frac{K}{\hbar} y^2 + \frac{Q}{\hbar} \ln (\frac{K}{\hbar}
y^2) \Bigr) \Bigr], \label{sing}
\end{equation}
so it has only the incoming flux toward the black hole singularity
as will be shown later.

We calculate the incoming (incident) flux at spatial infinity
\begin{eqnarray}
j_{in.} &=& {\rm Im} \Bigl[ \psi^* (y) i \frac{\hat{\pi}_y}{K}
\psi(y) \Bigr] \nonumber\\ &=& - \vert \tilde{D}_{I} \vert^2
\frac{Q \cosh\frac{\pi}{2\hbar}(Q - K)}{(\hbar K)^{1/2} \sinh
\frac{\pi}{\hbar} Q} e^{\frac{\pi}{2\hbar}(2Q + K)},\label{inc
flux}
\end{eqnarray}
where we used the relations $\vert \Gamma(1 + ix) \vert^2 =
\frac{\pi x}{\sinh \pi x}$ and  $\vert \Gamma(\frac{1}{2} + ix)
\vert^2 = \frac{\pi}{\cosh \pi x}$ \cite{abramowitz}. On the other
hand, the outgoing (reflected) flux at spatial infinity is
similarly found,
\begin{equation}
j_{ref.} = \vert \tilde{D}_{I} \vert^2 \frac{Q
\cosh\frac{\pi}{2\hbar} (Q + K)}{(\hbar K)^{1/2}
\sinh\frac{\pi}{\hbar} Q} e^{\frac{\pi}{2\hbar} K}. \label{ref
flux}
\end{equation}
The rate for black hole formation is the ratio of the transmitted
flux to the incident flux. From the flux conservation applied to
Eqs. (\ref{inc flux}) and (\ref{ref flux}) we obtain the
(transmission) rate for the black hole formation
\begin{eqnarray}
\frac{j_{tran.}}{j_{in.}} &=& 1 - \frac{j_{ref.}}{j_{in.}}
\nonumber\\ &=& 1- \frac{\cosh\frac{\pi}{2\hbar}(Q + K)
}{\cosh\frac{\pi}{2\hbar} (Q - K)} e^{- \frac{\pi}{\hbar} Q}.
\label{trans rate1}
\end{eqnarray}
Eq. (\ref{trans rate1}) determines the probability for the black
hole formation for all values of $c_0$.

In order to understand the classical limit of this solution, let
us consider the large $K$ limit. $K$ always enters Eqs.
(\ref{wav1}), (\ref{asym}), and (\ref{trans rate1}) in the
combination of $\frac{K}{\hbar}$, so the large $K$ limit is
equivalent to the small $\hbar$ limit, which is the semiclassical
limit. From the mass of black hole formed in the supercritical
case \cite{frolov}
\begin{equation}
M_{AH} = \frac{m_P^2}{\hbar} \Bigl( \frac{1}{2} r_{AH} \Bigr) =
\frac{m_P^2}{\hbar} \Bigl( \frac{\omega_c}{2} y_{AH} \Bigr)
\label{bh mass}
\end{equation}
we can express $\frac{K}{\hbar}$ as
\begin{equation}
\frac{K}{\hbar} = \frac{m_P^2}{\hbar^2} \Bigl(
\frac{\omega_c^2}{2} \Bigr) = \frac{4}{c_0^2}
\Bigl(\frac{M_{AH}}{m_P} \Bigr)^2, \label{k-bh}
\end{equation}
where $\omega_c$ denotes the cutoff of the $\omega$ variable. In
the supercritical case the influx from spatial infinity falls into
the black hole singularity and the black hole mass increases
infinitely for an infinite duration. Hence the large $K$ limit is
the large black hole mass limit. In this limit the rate for black
hole formation should be calculated separately for $c_0 > 1$ and
$c_0 < 1$. For $c_0 > 1$ Eq. (\ref{trans rate1}) becomes
asymptotically
\begin{equation}
\frac{j_{tran.}}{j_{in.}} = 1- e^{- \frac{\pi}{\hbar} K (c_0 -1)}.
\end{equation}
Therefore, the rate for black hole formation becomes unity for
$c_0 > 1 $, which implies that a black hole is formed from the
collapsing scalar field, as in the classical case \cite{frolov}.

In the case of $c_0 < 1$, we carefully rewrite Eq. (\ref{trans
rate1}) as
\begin{equation}
\frac{j_{tran.}}{j_{in.}} = \frac{e^{- \frac{\pi}{2\hbar}(K + Q)}
\sinh \frac{\pi}{\hbar} Q}{\cosh \frac{\pi}{2\hbar} (K - Q)}.
\label{trans rate2}
\end{equation}
The asymptotic value of Eq. (\ref{trans rate2})
\begin{equation}
\frac{j_{tran.}}{j_{in.}} = e^{- \frac{\pi}{\hbar} K (1 - c_0)},
\end{equation}
is the tunneling rate for quantum black hole formation in the
subcritical case \cite{bak}. The classical result is obtained in
the very large limit of $K$, in which the scalar field with $c_0 >
1$ evolves supercritically to form a black hole and with $c_0 <
1$, it bounces back without forming a black hole.

A few comments are in order. Firstly, the flux conservation is
valid at spatial infinity and the singularity. To show the
conservation we compute the transmission flux from spatial
infinity toward the singularity
\begin{eqnarray}
j_{tran.} &=& j_{in.} - j_{ref.} \nonumber\\ &=& \vert
\tilde{D}_{I} \vert^2 \frac{Qe^{\frac{\pi}{2\hbar}Q}}{(\hbar
K)^{1/2}},
\end{eqnarray}
which coincides with the flux near $y = 0$ obtained by a direct
computation using Eq. (\ref{sing}), as expected. Secondly, since
Eq. (\ref{wd eq}) is linear and $c_0$ is just a separation
parameter, one may construct a more general solution for the black
hole formation by superposing the solutions (\ref{asym}) with
different $c_0$
\begin{equation}
\Psi^{F}_{BH} (y, \phi) = \int d c_0 F (c_0) \psi_{BH} (y) \exp
\Bigl( \pm i \frac{K c_0}{\hbar^{1/2} m_P} \phi \Bigr), \label{sup
wav}
\end{equation}
where $F(c_0)$ is an arbitrary weighting factor. By recalling that
$(\mu_- - a_-)$ from Eq. (\ref{parameter}) and $(\mu_- - a_+^*)$
from Eq. (\ref{parameter2}) are independent of $c_0$, one still
has the factorized asymptotic form for Eq. (\ref{sup wav})
\begin{equation}
\Psi^{F}_{BH} (y, \phi) \cong F_{(-)} (\phi) \Bigl(i
\frac{K}{\hbar} y^2 \Bigr)^{\mu_{-} - a_{-}} e^{- \frac{i}{2}
\frac{K}{\hbar} y^2}+ F_{(+)} (\phi) \Bigl(i \frac{K}{\hbar} y^2
\Bigr)^{\mu_{-} - a_{+}^*} e^{ \frac{i}{2} \frac{K}{\hbar} y^2}.
\end{equation}
Here, $F_{(\pm)} (\phi)$, which are obtained by integrating over
$c_0$, may represent wave packets of plane waves, $e^{\pm i
\frac{K c_0}{\hbar^{1/2} m_P} \phi}$, in Eq. (\ref{fac wav}).

Finally, we explain the reason why the other wave functions for
Eq. (\ref{wd eq}) are irrelevant for the black hole formation. The
wave function (\ref{wav3}) has incident components from both
spatial infinity and the black hole singularity to the potential
barrier, so it does not satisfy the boundary condition for the
black hole formation. Another wave function is obtained by taking
a linear combination of Eqs. (\ref{wav1}) and (\ref{wav3}) which
describes the outgoing component alone at spatial infinity,
\begin{eqnarray}
\psi_{IV} (y) &=& \psi_I - \frac{\tilde{D}_I}{\tilde{D}_{III}}
e^{i \pi (a_- - a_+)} \frac{\Gamma (b_-) \Gamma
(a_{-}^*)}{\Gamma(b_+ ) \Gamma(a_{+}^*)} \psi_{III} \nonumber\\
&\cong& \Bigl[1 - e^{i \pi (a_{-} - a_{+})} \frac{\Gamma(a_-)
\Gamma(a_{-}^*)}{\Gamma(a_+) \Gamma(a_{+}^*)} \Bigr] \tilde{D}_{I}
\frac{\Gamma(b_{-})}{\Gamma( a_{-})} \Bigl(i \frac{K}{\hbar} y^2
\Bigr)^{\mu_{-} - a_{+}^*} e^{ \frac{i}{2} \frac{K}{\hbar} y^2} .
\label{wav4}
\end{eqnarray}
Since the net flux is the difference between those of the incoming
and outgoing components, it is easy to see that the one-parameter
($\theta$) family of wave functions,
\begin{equation}
\psi_{\theta} (y) = \psi_I + \frac{e^{i \theta} - 1}{1 - e^{i \pi
(a_- - a_+)} \frac{\Gamma(a_-) \Gamma(a_{-}^*)}{\Gamma(a_+)
\Gamma(a_{+}^*)}} \psi_{IV} \label{theta sol}
\end{equation}
has the same rate as Eq. (\ref{trans rate1}). However, the
branches $\psi_{I}$ and $\psi_{IV}$ of the wave function
(\ref{theta sol}) have the incident components both from spatial
infinity and from the black hole singularity to the potential
barrier, respectively. We also note that the wave function
$\psi_{IV}^*$, the complex conjugate of Eq. (\ref{wav4}), is also
a solution to the linear real equation (\ref{wd eq}) and is purely
incident from spatial infinity. However, near $y = 0$,
$\psi_{IV}^*$ consists of two branches of wave functions:
$\psi_{I}^*$ having an outgoing flux and $\psi_{III}^*$ having an
incoming flux. Thus $\psi_{IV}^*$ describes the incident waves
from both spatial infinity and the black hole singularity to the
potential barrier. Hence these wave functions, $\psi_{IV}^*$ and
$\psi_{\theta} (\theta \neq 0)$, do not satisfy in a strict sense
the boundary condition for the black hole formation.

\section{Semiclassical Limit and Apparent Horizon}

In order to regain the semiclassical picture from our quantum wave
function we use our recently proposed method \cite{bak2}. For this
we first consider the Wheeler-DeWitt equation with nonvanishing
energy parameter $E$ given by
\begin{equation}
\Biggl[ \frac{\hbar^2}{2K} \frac{\partial^2}{\partial y^2} -
\frac{\hbar m_P^2}{2Ky^2} \frac{\partial^2}{\partial \phi^2} - K
\Bigl(1 - \frac{y^2}{2} \Bigr) \Biggr] \Psi_E(y, \phi) = E \Psi_E
(y, \phi), \label{e-wd}
\end{equation}
where $E$ has the dimension of $\hbar$. The solution corresponding
to the black hole formation is
\begin{equation}
\psi_E (y) = D  \Biggl[ \exp \Bigl(- \frac{i}{2} \frac{K}{\hbar}
y^2 \Bigr) \Biggr] \Bigl(\frac{K}{\hbar} y^2 \Bigr)^{\mu_{-}} M
(a_{E}, b_{-}, i \frac{K}{\hbar} y^2), \label{e-wav}
\end{equation}
where $\mu_{-}$ and $b_{-}$ are the same as Eq. (\ref{parameter}),
while
\begin{equation}
a_E = \frac{1}{2} - \frac{i}{2\hbar} \Bigl( Q + K + E \Bigr).
\end{equation}
Note that the parameter $E$ appears indirectly only through $a_E$.
The Hamilton-Jacobi characteristic function $W_E (y)$ is given as
\begin{eqnarray}
W_E (y) &=& \hbar {\rm Im} \ln \psi_E (y) \nonumber\\ &=& -
\frac{1}{2} K y^2 - \frac{1}{2} Q \ln (\frac{K}{\hbar}y^2) + \hbar
{\rm Im} M(a_E, b_-, i\frac{K}{\hbar} y^2) + {\rm constant}.
\label{char fn}
\end{eqnarray}
From Eq. (\ref{char fn}) we recover the evolution equation of the
gravitational collapse as
\begin{eqnarray}
\tau + \beta &=&  \frac{\partial}{\partial E} W_E (y) \Bigl|_{E =
0} \nonumber\\ &=& \hbar \frac{\partial}{\partial E} \Bigl[{\rm
Im} \ln M (a_E, b_-, i \frac{K}{\hbar} y^2) \Bigr] \Bigl|_{E = 0},
\label{hj eq}
\end{eqnarray}
where $\beta$ is a constant to be determined by an initial
condition.

Before turning to the evolution equation (\ref{hj eq}) in the
region where quantum effects are important, we compare it with the
classical equation of motion in the regions where classical
effects are dominant. The classical solution of the evolution
equation
\begin{equation}
\frac{\dot{y}^2}{2} + 1 - \frac{y^2}{2} - \frac{c_0^2}{2 y^2} = 0,
\label{cl eq}
\end{equation}
is given by
\begin{eqnarray}
\tau + \beta &=& \pm \int dy \frac{1}{\sqrt{y^2 +
\frac{c_0^2}{y^2} - 2}} \nonumber\\ &=& \pm \frac{1}{2} \ln
\Bigl(y^2 - 1 + \sqrt{y^4 - 2 y^2 + c_0^2} \Bigr). \label{cl sol}
\end{eqnarray}
Firstly, at spatial infinity $(y \rightarrow \infty)$ we have the
asymptotic expression
\begin{equation}
\tau + \beta = \pm \frac{1}{2} \ln y^2 + {\rm constant}.
\end{equation}
We compare this classical evolution with the quantum one. We
compute the characteristic function for the incoming and out-going
components, separately, using Eq. (\ref{hj eq})
\begin{eqnarray}
\tau + \beta &=& \hbar {\rm Im} \Biggl[\frac{-i}{2M}
\frac{\partial}{\partial a} M (a, b_-, i\frac{K}{\hbar} y^2)
\Bigl|_{a = a_-, y \rightarrow \infty}\Biggr] \nonumber\\ &=& \pm
\frac{1}{2} \ln y^2,
\end{eqnarray}
where the upper (lower) sign comes from the outgoing (incoming)
component in Eq. (\ref{asym}). Hence in the asymptotic region the
Hamilton-Jacobi solution and the classical limit give the same
evolution of the collapsing process.

Secondly, near the singularity $(y \rightarrow 0)$ the allowed
motion is toward the singularity $y = 0$. So the classical
solution is given by the lower sign in Eq. (\ref{cl sol}) and has
the asymptotic form
\begin{eqnarray}
\tau + \beta &=& - \frac{1}{2} \ln \Bigl[c_0 - 1 + y^2 \Bigl(1 -
\frac{1}{c_0} \Bigr) \Bigr] \nonumber\\ &=&  \frac{y^2}{2c_0} +
{\rm constant}.
\end{eqnarray}
On the other hand, the wave function (\ref{e-wav}) depends on $E$
only through the confluent hypergeometric function $M (a_E, b_-, i
\frac{Ky^2}{\hbar})$, and the semiclassical limit of the quantum
solution is
\begin{eqnarray}
\tau + \beta &=& \hbar \frac{\partial}{\partial E}  \Bigl[{\rm Im}
\ln M (a_E, b_-, i \frac{K}{\hbar} y^2) \Bigr] \nonumber\\ &=&
\frac{y^2}{2c_0}.
\end{eqnarray}
Thus, in the semiclassical limit $(\hbar \rightarrow 0$, which is
equivalent to $K \rightarrow \infty)$ the quantum wave function
has the correct limit to the Hamilton-Jacobi solution both in the
asymptotic region and near the singularity.

In the intermediate region between the asymptotic region and the
singularity it is most interesting to see how the apparent horizon
and black hole mass are affected by quantum effects. We note that
the position of the apparent horizon is given by the trapped
surface $\bigl(\nabla r \bigr)^2 = 0$, which translates to
$\dot{y}^2 - y^2 = 0$. In the classical case the apparent horizon
is determined from the equation of motion,
\begin{equation}
\frac{d W_c}{d y} = - \sqrt{K^2 \Bigl(y^2 + \frac{c_0^2}{y^2} - 2
\Bigr)} = K \dot{y} = - K y, \label{cl-ah eq}
\end{equation}
which has the solution
\begin{equation}
y_{AH}^2 = \frac{c_0^2}{2}. \label{cl-ah}
\end{equation}
On the other hand, in the quantum case we use the quantum
counterpart Eq. (\ref{char fn}) to Eq. (\ref{cl-ah eq}),
\begin{equation}
\frac{d W_E}{dy} \Bigl|_{E = 0} =  K \dot{y} = - K y. \label{q-ah
eq}
\end{equation}
Since $W|_{E =0} = \hbar {\rm Im} \ln \psi$, Eq. (\ref{q-ah eq})
becomes
\begin{eqnarray}
-K y &=& \hbar \frac{d}{dy} \Bigl( {\rm Im} \ln \psi \Bigr)
\nonumber\\ &=& - Ky - \frac{Q}{y} + \hbar {\rm Im} \Biggl[
\frac{d}{dy} \ln M \Biggr],
\end{eqnarray}
which can be rewritten as
\begin{equation}
\frac{Q}{2K} = {\hbar}y^2 \mbox{Re}\Biggl[\frac{a}{b}\frac{M(a+1,b+1,z)}{M(a,b,z)}\Biggr].
\end{equation}
In the semiclassical limit ($K/\hbar\rightarrow \infty$) we can evaluate the
apparent horizon ($y^2_{AH}$) using the steepest descent method (see the Appendix
for details).  We find the apparent horizon up to the first order as
\begin{equation}\label{appr}
y^2_{AH} = \frac{{c_0}^2 }{2} + \frac{{\hbar}^2
}{K^2}\frac{3-\frac{64}{{c_0}^2}
+\frac{80}{{c_0}^4}}{8} + O(\frac{\hbar^3}{K^3}),
\end{equation}
where the first term in the right hand side is the classical
result and the second term is the first order quantum correction.
We checked this with numerical calculation for large $K/\hbar (
=40)$. They are
\begin{eqnarray}
y^2_{AH} \Big|_{\rm numerical} &=& 1.9616 (c_0 = 2) ;~ 0.5000 (c_0
= 1); ~0.2569 (c_0 = \frac{1}{\sqrt{2}}), \nonumber\\
y^2_{AH}\Big|_ {\rm Eq.(55)} &=& 1.9994 (c_0 = 2);~ 0.5001 (c_0 =
1); ~0.2652 (c_0 = \frac{1}{\sqrt{2}}) .
\end{eqnarray}
Even in the critical case ($c_0=1$) the apparent horizon deviates only
slightly from the classical one, and the first order correction is
quite good.
As far as the apparent horizon is concerned the critical case
seems to show no special behavior compared to the super and
subcritical cases.  For small $K/\hbar$ there is a large quantum correction, and
the approximation~(\ref{appr}) is not valid.  For example we calculated
numerically when $K/\hbar=1$, and the deviation from the classical value is
substantial as
\begin{equation}
y^2_{AH} \Big|_{\rm numerical} - y^2_{AH} \Big|_{\rm classical} =
-0.412 (c_0 = 2) ;~ 0.127 (c_0 = 1); ~0.128 (c_0 =
\frac{1}{\sqrt{2}}).
\end{equation}
However, there is no qualitative change in the position of the horizon.

\section{Discussion}

In this paper we have studied the black hole formation by a
self-similarly collapsing massless scalar field. The analytic wave
functions for the black hole formation were found in terms of the
confluent hypergeometric function. By evaluating the incoming and
outgoing flux at spatial infinity we were able to compute the rate
for quantum black hole formation, which agrees with our previous
result in the subcritical case. To find other quantum effects we
used the characteristic function defined by the imaginary part of
the wave function and recovered the evolution of black hole
formation in time with quantum effects taken into account. We now
compare our results with those from the classical solution, and
the wave functions by Tomimatsu, and discuss the physical
implications.

Firstly, in the classical case \cite{frolov} the evolution of the
gravitational collapse is governed by Eq. (\ref{cl eq}) and the
solution is given by Eq. (\ref{cl sol}). In the supercritical case
the incoming component, the lower sign of Eq. (\ref{cl sol}),
falls into the singularity and any part of it is not reflected
back to spatial infinity. On the other hand, in the quantum case
the wave function (\ref{wav1}) for black hole formation has both
the incoming and outgoing components (\ref{asym}). In particular,
in the supercritical case it has a small fraction of outgoing
component. In the subcritical case the black hole formation is
allowed by quantum mechanical tunneling.

Secondly, among all the wave functions (\ref{wav1}), (\ref{wav2})
and (\ref{wav3}) for the Wheeler-DeWitt equation, Eq. (\ref{wav1})
has the desired semiclassical limit for black hole formation. On
the other hand, Eq. (\ref{wav4}) represents a wave function with
outgoing component alone. Though the physical meaning of this wave
function is not clear, it may be interpreted as an amplitude for
the black hole decay or white hole creation, which requires a
further study.

Thirdly, as mentioned in Sec. IV, the apparent horizon in the
quantum case deviates from the classical one only slightly. Thus
there appears no qualitative change from the classical to the
quantum treatment of the apparent horizon.

Finally, we compare our wave functions with those by Tomimatsu who
found the wave functions for the quantum black hole formation and
decay \cite{tomimatsu}. The difference is that he used a different
coordinate system from ours and quantized both the classical
constraint equation and the Hamiltonian from the reduced action,
whereas we adopted directly the ADM formulation, obtained the
Hamiltonian constraint and quantized the constraint according to
the Dirac quantization method. In contrast to his wave functions
diverging at the origin in both the supercritical and subcritical
cases, our wave functions are always regular at the origin. The
regularity of wave function for black hole seems to suggest that
quantum gravity effects may cure some singularity problems in
classical gravity.

\begin{acknowledgements}
We would like to thank Dr. J. Y. Ji for numerical calculations,
and P. R. Brady, B. Carr, M. W. Choptuik, V. Frolov, A. Hosoya, D.
N. Page, and A. Tomimatsu for many useful discussions and comments
during YKIS' 99. Also we would like to appreciate the warm
hospitality of CTP of Seoul National University where this paper
was completed. This work was supported in parts by BSRI Program
under BSRI 98-015-D00054,98-015-D00061, 98-015-D00129. DB was
supported in part by KOSEF Interdisciplinary Research Grant
98-07-02-07-01-5, SPK by KRF under Grant No. 98-001-D00364, and
JHY by KOSEF under Grant No. 98-07-02-02-01-3.
\end{acknowledgements}

\appendix
\section*{Useful Formulas for the Confluent Hypergeometric
Functions}

The asymptotic form of the confluent hypergeometric function
\cite{abramowitz} used in this paper is
\begin{equation}
\frac{M(a, b, z)}{\Gamma(b)} = \frac{e^{i \pi a} z^{-a}
}{\Gamma(b-a)} \Bigl[1 + \frac{e^{z} z^{a-b}}{\Gamma (a)} \Bigr],
\end{equation}
where $ - \frac{\pi}{2} < {\rm arg} ~ z < \frac{3\pi}{2}$.

To obtain the characteristic function valid at intermediate region
we use the integral formula for the confluent hypergeometric
function \cite{abramowitz}
\begin{equation}
\frac{\Gamma(b - a) \Gamma (a)}{\Gamma (b)} M (a, b, z) =
\int_{0}^{1} dt e^{z t} t^{a-1} (1 - t)^{b-a-1},\label{form 1}
\end{equation}
which is valid for ${\rm Re}~ b > {\rm Re}~ a > 0$. In this paper
$z = i \frac{K}{\hbar} y^2$ and ${\rm Re} ~b = 1 > {\rm Re}~ a =
\frac{1}{2} > 0$.
In order to evaluate $M(a,b,z)$ in the large $K$ limit we evaluate it
with the steepest descent method.  For this we define
\begin{equation}
\frac{iK}{\hbar}f(t) = tz+(a-1)\ln (t) +(b-a-1)\ln (1-t).
\end{equation}
We find the relevant root of $\frac{df}{dt}=0$ as
\begin{equation}
t_{-} = \frac{c+y^2-(y^4+c^2-2y^2)^{1/2}}{2y^2},
\end{equation}
where $c=({c_0}^2-\frac{\hbar^2}{4K^2})^{1/2}-i\frac{\hbar}{K}$.
We evaluate the integral
\begin{equation}
\int dt~ \exp \Bigl(i\frac{K}{\hbar}f(t)\Bigr) = \Biggl[ \exp
\Bigl(i\frac{K}{\hbar}f(t_-)\Bigr) \Biggr]
\Bigl(\frac{2\pi\hbar}{-iKf^{''}(t_{-})}\Bigr)^{1/2}
\Bigl(1+O(\frac{\hbar}{K}) \Bigr).
\end{equation}
After straightforward calculations we obtain
\begin{equation}
y^2_{AH} = \frac{{c_0}^2}{2}\Biggl[1+\frac{\hbar^2}{K^2}
\frac{3-\frac{64}{{c_0}^2 }+ \frac{80}{{c_0}^4}}{4{c_0}^2}
+O(\frac{{\hbar}^2}{K^3})\Biggr],
\end{equation}
which is in good agreement with the numerical calculations in the
large $K$ limit as shown at the end of Sec. IV.

\end{document}